\documentclass[preprint2]{aastex}

\newcommand{\vtau}{\mbox{V471 Tau}}

\newcommand{\asca}{\mbox{\em ASCA}}
\newcommand{\chandra}{\mbox{\em Chandra}}
\newcommand{\xmm}{\mbox{\em XMM-Newton}}
\newcommand{\exosat}{\mbox{\em EXOSAT}}
\newcommand{\rosat}{\mbox{\em ROSAT}}
\newcommand{\euve}{\mbox{\em EUVE}}
\newcommand{\hst}{\mbox{\em HST}}

\newcommand{\iue}{\mbox{\em IUE}}

\newcommand{\dex}[1]{\hbox{$\times\hbox{10}^{#1}$}}

\newcommand{\etal}{\mbox{et\ al.\ }}

\slugcomment{Submitted to the Astrophysical Journal}

\shorttitle{Coronal abundances of V471 Tau}
\shortauthors{M.~Still and G.~Hussain}

\begin{document}

\title{Coronal element abundances of the post-common envelope 
binary V471~Tauri with {\em ASCA}}

\author{Martin Still\altaffilmark{1}}
\affil{NASA/Goddard Space Flight Center, Code 662, Greenbelt, MD~20771}

\author{Gaitee Hussain}
\affil{Harvard-Smithsonian Center for Astrophysics, 60 Garden Street, 
Cambridge, MA 02138}

\altaffiltext{1}{Universities Space Research Association}

\begin{abstract} 

We report on \asca\   observations of the coronally  active  companion
star  in  the   post-common   envelope  binary   \vtau.   Evolutionary
calculations indicate that there  should be no peculiar  abundances on
the companion star resulting from the common  envelope epoch.  Indeed,
we  find no evidence  for peculiar  abundances, although uncertainties
are high. We find that a single-temperature plasma  model does not fit
the data.  Two-temperature  models with decoupled elemental abundances
suggest that  Fe is underabundant compared  to the Hyades photospheric
mean.  In the absence of a measurement of photospheric Ne abundance in
the  cluster,  we find  Ne is   overabundant   compared to   the solar
photospheric    value.  This is    indicative  of  the inverse   first
ionization   potential  effect.   Differences   between   coronal  and
photospheric abundances are believed  to result from the fractionation
of ionized and neutral material in the upper atmosphere of the star.

\end{abstract}


\section{Introduction}
\label{sec:introduction}

Solar  observations  have    indicated that  coronal  abundances   are
photospheric for  high-FIP (First  Ionization Potential) elements, but
that the coronal plasma is over abundant  in low-FIP elements (Feldman
1992).  Since low-FIP elements  are ionized in the solar chromosphere,
whereas high-FIP elements  are not (Geiss 1992),  it is  believed that
some mechanism  separates ions  from neutrals within  the chromosphere
and transports  them to the corona.  H\'{e}noux 1995 provides a review
of suggested mechanisms.

CCD X-ray detectors  and   proportional  counting  spectrographs  have
revealed multi-temperature stellar coronae with a variety of abundance
distributions  over ionization  potential.    Some show  a  FIP effect
similar to  the   sun, others  show  the  opposite  trend -  abundance
decreasing  with  FIP relative  to   the solar  photosphere, i.e.   an
inverse  FIP effect.   However,   because  of  the  moderate  spectral
resolutions available, combined with  the confusion of  line blending,
much has been made of the possible ambiguity in abundance measurements
using spectral models of collisionally-excited plasmas (Drake 1996).

The grating spectrographs   onboard \euve\ (Drake, Laming and   Widing
1996),  and  recently launched  on the \xmm\  and \chandra\ satellites
(Brinkman \etal  2001; Audard \etal 2001)  have confirmed  the general
results  of  lower-resolution spectroscopy.  Consequently, despite the
poorer resolution compared to grating  cameras and the possibility  of
line   confusion, confidence     in  CCD  and   proportional   counter
measurements has   increased.   In  this   paper   we  present  \asca\
spectroscopy of the   K star in the   close binary \vtau,  in order to
measure  its  coronal  content and  contrast   this with  photospheric
abundances.

\vtau\ is a member of the Hyades open cluster. Photospheric  abundances
of the  Hyades members have been  measured by Cayrel  \etal (1985) and
Varenne \&  Monier  (1999) and  some specific abundances  of \vtau\ by
Mart\'{i}n \etal (1997). Accepting model assumptions,  we show in this
paper  that   coronal Fe  is   under-abundant  relative  to the   mean
photospheric Hyades content and Ne is over-abundant, at least relative
to the  solar photosphere.  This may  be  indicating that there  is an
inverse-FIP effect in the  upper atmosphere of  the star.

\vtau\ is a 12.5 hour eclipsing binary with a white dwarf and 
tidally-locked K2 companion (Nelson and Young 1970).  At a distance of
47 pc (Werner \& Rauch 1997), it is the  closest object which has been
through a recent common  envelope phase of  evolution.  During the red
giant phase of the white dwarf progenitor's  life, the envelope of the
giant was large  enough to contain the  orbit of the K star companion.
While the two stars shared a common envelope, tides, friction and mass
loss  created a  significantly  smaller binary,  with  a short orbital
period (Pacz\'{y}nksi 1976).  The envelope of the giant has since been
ejected, but  the  binary still loses  angular  momentum  through wind
braking and gravitational radiation  (Taam 1983).  \vtau\ is therefore
one  of the best  candidates for  a pre-cataclysmic variable (pre-CV).
CVs occur when  the orbital period of  a white dwarf--red dwarf binary
becomes short enough  for   the main-sequence companion to   fill it's
Roche lobe.  Quasi-persistent accretion will  then occur through Roche
lobe overflow.  These objects are the  sources of dwarf nova outbursts
and classical nova  eruptions   that enrich the galactic  ISM  (Warner
1995).

Soft X-rays from \vtau\ were discovered by HEAO  1 (van Buren, Charles
\& Mason 1980) and it  was detected as  a  flaring source by the  {\em
Einstein}  Observatory (Young \etal 1983).  This  suggested that the K
star was  the source  of X-rays,  but by  determining the  white dwarf
temperature to be  35,000 K using the  {\it IUE} satellite, Guinan  \&
Sion (1984) showed that  the compact object  may also be a significant
X-ray emitter. Jensen \etal\ (1986) used  \exosat\ to find white dwarf
eclipses, pulsations,  orbital  dips and   emission from both  stellar
components.  Using the simultaneous EUV/X-ray   data from the  \rosat\
all-sky survey,    Barstow  \etal (1992) determined   that   the 555 s
pulsations were the result of accretion caps on the white dwarf rather
than g-mode oscillations. The interaction  of rotating magnetic fields
from the  white dwarf  and K  star  result in non-thermal  emission at
radio  energies between   the  two  objects  (Patterson, Caillault  \&
Skillman  1993; Lim,  White \& Cully  1996; Nicholls  \& Storey 1999).
Possibly it  is  this magnetically ejected  material  that is accreted
upon the white dwarf.  Wheatley   (1998) employed \rosat\ to  separate
the K star from the white dwarf spectrally and  pointed out that the K
star flux was  brighter than the average  Hyades member but similar to
the  younger, faster  rotators   of the  Pleiades.  Therefore  coronal
luminosity in this source is probably related  to rotation rate rather
than age. The white dwarf is softer spectrally  than the \asca\ energy
band, therefore it is not a factor during the  analysis of the current
data.

\begin{figure*}
\begin{picture}(100,0)(10,20) 
\put(0,0){\includegraphics{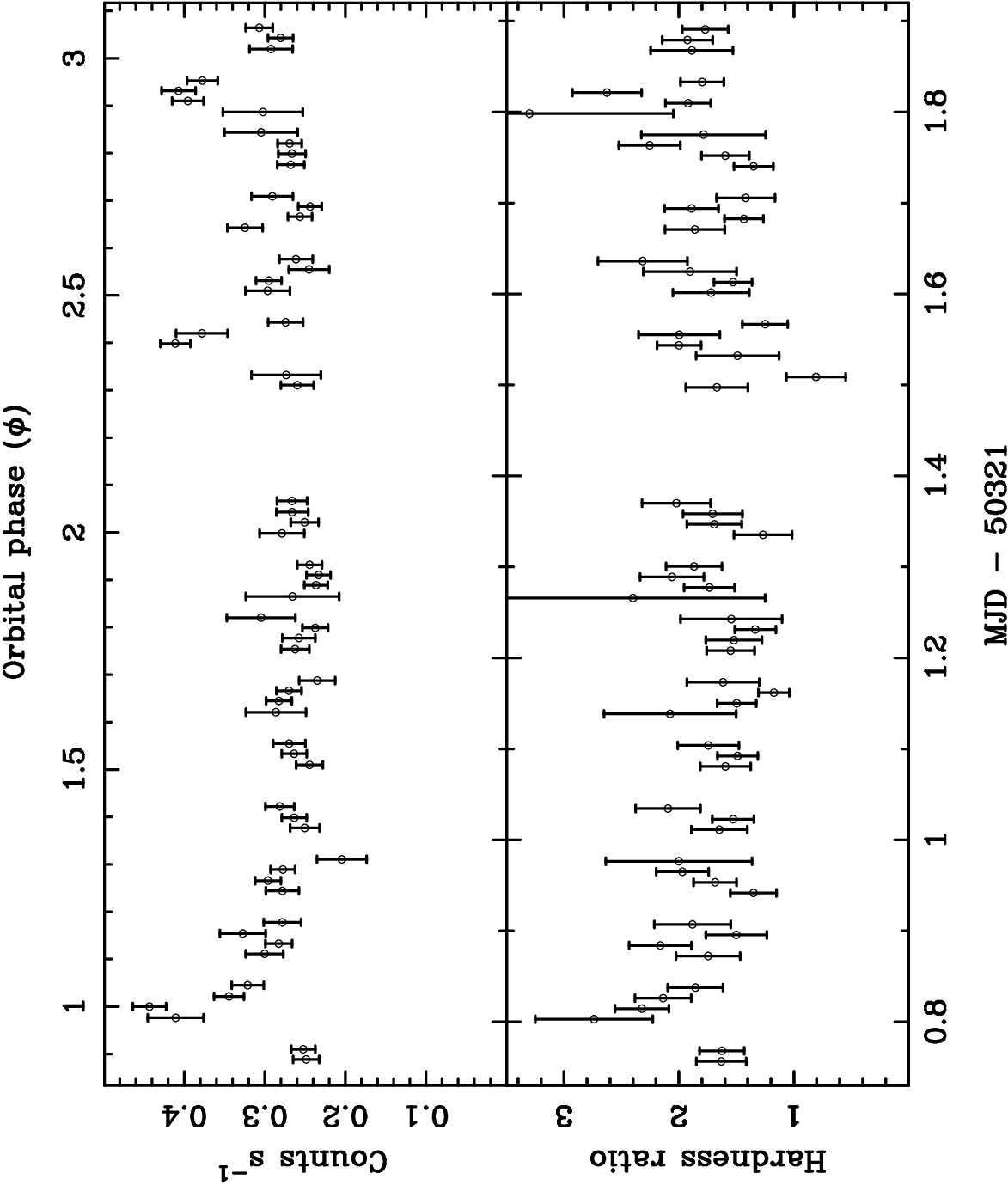}}
\noindent
\end{picture}
\vspace{86mm}
\figcaption[lc.ps]{Top: the combined SIS and GIS 0.3--10 keV light 
curve  for  \vtau.  Bottom: the hardness  ratio  between  energy bands
0.3--1.0 keV  and 1.0--10.0 keV.  The  orbital phase  defined from the
ephemeris of  Guinan \& Ribas (2001)  is also provided. No emission is
expected  from   the white  dwarf    at energies   greater  then   0.3
keV. \label{fig:lc}}
\end{figure*}

\section{Observations}
\label{sec:observations}

The Advanced Satellite  for Cosmology and Astrophysics (\asca; Tanaka,
Inoue \& Holt 1994) comprised four identical telescopes of nested thin
foil reflectors (Serlemitsos \etal 1995). Four cameras were mounted at
the focal planes.  Two had Solid state Imaging Spectrometers (SIS0 and
SIS1; Burke  \etal 1991), and  two  carried Gas  Imaging Spectrometers
(GIS2 and GIS3) which were scintillation proportional counters (Ohashi
\etal 1991).   The SIS cameras have   an energy range   0.4--10 keV, a
spectral resolution of  $E/\Delta E$  = 20 at  2  keV and a 22  arcmin
field of view. The GIS instruments have an energy range 0.7--10 keV, a
spectral resolution of  $E/\Delta E$ = 7.5  at 2 keV  and  a 50 arcmin
field of view. 

Between MJD  50321.75 and MJD  50322.90  (1996 Aug 26-7),  \asca\  was
pointed at  \vtau\  for a total   exposure  time of  50 ksec,  over 17
near-contiguous orbits.  The  ASCA sequence number of the observations
is  24032000.  The  SIS cameras operated  in  FAINT mode and  the data
converted  to BRIGHT mode,  while GIS data was  taken in  PH mode with
normal bit assignments.

\section{Analysis}
\label{sec:analysis}

Observations were   screened  using standard criteria   and the FTOOLS
software  package v5.1. Events  were filtered based on the satellite's
proximity to South Atlantic Anomaly  passages, Earth elevation  angle,
pointing  stability and rigidity  of  the geomagnetic field.  Hot  and
flickering   pixels were removed from  the  SIS events and we rejected
particle events from the GIS observations.  SIS data were additionally
screened for  proximity to the  bright Earth-limb, time since the last
passage over the day-night terminator,  telemetry saturation and event
rate.   Only well-calibrated events,  with grades  0,  2, 3  or 4 were
retained. GIS data    were additionally screened spatially  to  remove
background  ring and calibration  source and further intervals of high
particle background were clipped.  A total of  41 ksec (SIS0), 44 ksec
(SIS1), 48 ksec (GIS2) and 48 ksec (GIS3) of  good data remained.  The
hot white dwarf component is  softer spectrally than the \asca\ energy
band (Wheatley 1998), we find no evidence for it  in the current data,
and therefore we do not consider it in the following analysis.

\subsection{Photometry}
\label{sec:photometry}

SIS   event rates  were    determined  within two  spatially-distinct,
circular region masks of diameter 8 arcmin and 5 arcmin for the source
and  background, respectively. Source  events  were extracted from the
GIS tables   within a  circular region  12  arcmin  in diameter  and a
background annulus  12--24 arcmin from  the  target.  Background rates
were  normalized  by  the appropriate  ratio  of extraction  areas and
subtracted  from  the source events.  The   total SIS+GIS source event
rate during the observation is provided in Fig.~\ref{fig:lc}, averaged
without weights  into 1 ksec bins.  We also provide the hardness ratio
in the bands 0.3--1.0 keV and 1.0--10.0 keV.

\begin{figure*}
\begin{picture}(100,0)(10,20) 
\put(0,0){\includegraphics{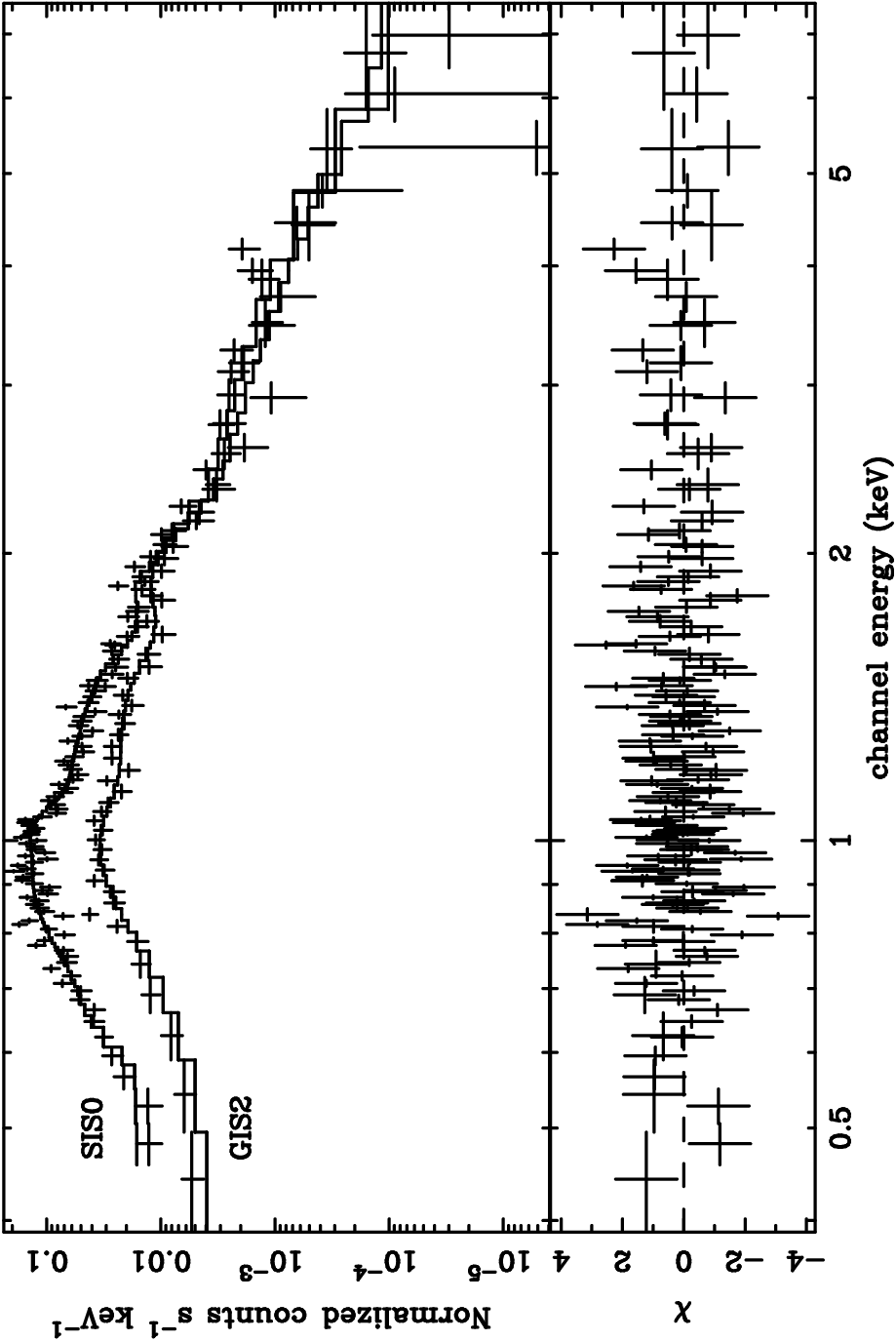}}
\noindent
\end{picture}
\vspace{84mm}
\figcaption[spectrum.ps]{SIS0  and GIS2 spectra of \vtau\  with best 
fit.   The model  is   an  absorbed, optically thin,  two  temperature
thermal    plasma     corresponding    to   Model     3      in  Table
\ref{tab:abundance}. \label{fig:spectrum}}
\end{figure*}

Phases  for each time bin  were calculated using the orbital ephemeris
of Guinan \& Ribas (2001), converted to Modified Julian Date.
\begin{equation}
T = \hbox{MJD}~40\,609.564056(11) + 0.521183398(26)\,E
\end{equation}
where $T$ is the average MJD of each bin and  $E$ is the cycle number.
The orbital  phase, $\phi$, in Fig.~\ref{fig:lc}   corresponds to $E -
18\,624$.     Quantities  in  parentheses   correspond   to  $1\sigma$
uncertainties.  

Count rate is approximately constant, except for three flares, each of
duration a few ksec at orbital phases  $\phi \simeq$ 1.0, 2.4 and 2.9.
Statistically, the shape of the  flares are not well-defined and there
is  only marginal evidence for  spectral  hardening. Flares of similar
duration and amplitude  are common to   the rapidly rotating  stars in
tidally-locked  binaries,  e.g., the   RS  CVn systems   (Franciosini,
Pallavicini  \&  Tagliaferri 2003),  and   to  single M  dwarf   stars
(Tsikoudi \&  Kellett 2000).   The observed   flares do not  occur  at
preferred orbital phases.

\vtau\ is relatively faint compared to most well-observed  coronal 
sources (e.g.\ Brinkman \etal 2001, Audard \etal 2001), so in order to
maximize  counting statistics  during  spectral  fitting,  we do   not
separate flare events from the quiescent data.

\begin{figure*}
\begin{picture}(100,0)(10,20) 
\put(0,0){\includegraphics{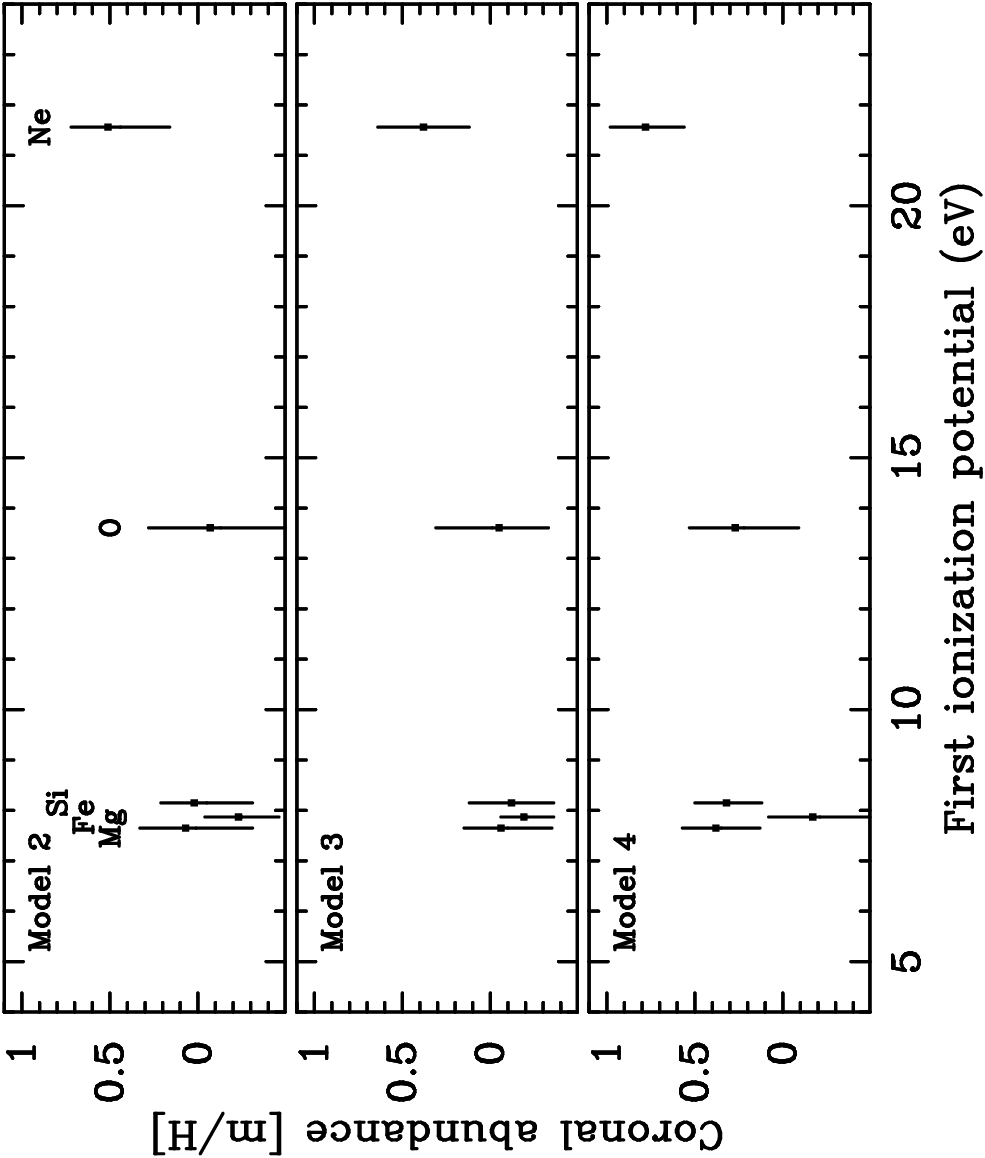}}
\noindent
\end{picture}
\vspace{90mm}
\figcaption[fip.ps]{Coronal elemental abundances from Models 2, 3 and 4, 
showing a  relative overabundance of Ne in  each case.  Error bars are
90 percent confidence ranges. \label{fig:fip}}
\end{figure*}

\subsection{Spectroscopy}
\label{sec:spectroscopy}

\begin{deluxetable}{cccccc}
\footnotesize
\tablecaption{Spectral fit parameters for four thermal plasma models. 
Abundances are provided as the  logarithmic value of the ratio between
coronal and solar  photospheric values, relative  to H. The abundances
of some  of the elements, C, N,  Al, S, Ar, Ca  and  Ni, were not well
constrained by the data.   Models are  identical except for  different
assumed abundances for the elements above.  Best fit abundances in the
\asca\  band are provided first,  with  the temperatures and  emission
measures,  assuming   a   two-temperature   emission region,   second.
Uncertainties are 90 percent confidence  limits for each quantity, All
fit parameters were free  to vary during the uncertainty calculations.
The $\chi^2$ of each fit and number of degrees of freedom are provided
at the bottom.\label{tab:abundance}}
\tablewidth{0pt}
\startdata
\hline\hline
Element & FIP (eV) & 
Model 1\tablenotemark{a} & 
Model 2\tablenotemark{b} & 
Model 3\tablenotemark{c} & 
Model 4\tablenotemark{d} \\
\hline
\vspace{-3.5mm}\\
\mbox{[Mg/H]} &
\hspace{.5em}7.65 &
$-0.23^{+0.12}_{-0.06}$ & 
 $0.07^{+0.38}_{-0.25}$ & 
$-0.06^{+0.29}_{-0.21}$ & 
 $0.38^{+0.25}_{-0.18}$ \\
\vspace{-3.5mm}\\
\mbox{[Fe/H]} &  
\hspace{.5em}7.87 & 
$-0.23^{+0.12}_{-0.06}$ & 
$-0.17^{+0.29}_{-0.12}$ &
$-0.19^{+0.17}_{-0.13}$ &
$-0.17^{+0.33}_{-0.25}$ \\
\vspace{-3.5mm}\\
\mbox{[Si/H]} &  
\hspace{.5em}8.15 & 
$-0.23^{+0.12}_{-0.06}$ & 
 $0.02^{+0.33}_{-0.19}$ &
$-0.12^{+0.24}_{-0.23}$ &
 $0.32^{+0.20}_{-0.17}$ \\
\vspace{-3.5mm}\\
\mbox{[O/H]}\hspace{.25em} & 
13.61 & 
$-0.23^{+0.12}_{-0.06}$ & 
$-0.07^{+0.47}_{-0.35}$ &
$-0.05^{+0.28}_{-0.35}$ &
 $0.27^{+0.36}_{-0.25}$ \\
\vspace{-3.5mm}\\
\mbox{[Ne/H]} & 
21.56 & 
$-0.23^{+0.12}_{-0.06}$ & 
 $0.51^{+0.35}_{-0.20}$ &
 $0.38^{+0.26}_{-0.25}$ &
 $0.78^{+0.22}_{-0.19}$ \\
\vspace{-4.mm}\\
\hline
& & & & & \\

\hline\hline
Parameter & &
Model 1\tablenotemark{a} & 
Model 2\tablenotemark{b} & 
Model 3\tablenotemark{c} & 
Model 4\tablenotemark{d} \\
\hline
\vspace{-3.5mm}\\
$kT_1$\tablenotemark{e,f} & &
$0.78^{+0.03}_{-0.02}$ &
$0.67^{+0.07}_{-0.07}$ & 
$0.71^{+0.06}_{-0.12}$ &
$0.67^{+0.08}_{-0.07}$ \\
\vspace{-3.5mm}\\
$EM_1$\tablenotemark{g}\hspace{.5em} & &
$3.05^{+0.82}_{-0.88}$ &
$2.02^{+1.13}_{-1.02}$ & 
$2.30^{+1.24}_{-0.91}$ &
$1.08^{+0.47}_{-0.46}$ \\
\vspace{-3.5mm}\\
$kT_2$\tablenotemark{e,f} & &
$1.93^{+0.18}_{-0.16}$ &
$2.51^{+0.49}_{-0.37}$ & 
$2.45^{+0.37}_{-0.38}$ &
$2.57^{+0.48}_{-0.24}$ \\
\vspace{-3.5mm}\\
$EM_2$\tablenotemark{g}\hspace{.5em} & &
$3.00^{+0.32}_{-0.37}$ &
$1.99^{+0.51}_{-0.77}$ & 
$2.60^{+0.85}_{-0.04}$ &
$1.26^{+0.20}_{-0.24}$ \\
\vspace{-4.mm}\\
\hline
\vspace{-3.5mm}\\
\hspace{1.5em}$\chi^2$(dof) & &
414(341) &
380(336) &
377(335) &
383(336) \\
\vspace{-4.7mm}\\
\enddata
\tablenotetext{a}{Metal abundances all equal.}
\tablenotetext{b}{Metal abundances equal to Fe, except Mg, Si, O 
and Ne.}
\tablenotetext{c}{Metal abundances equal, except Mg, Fe, 
Si, O and Ne.}
\tablenotetext{d}{Metal abundances equal to the Hyades photospheric 
mean, except Mg, Fe, Si, O and Ne.}
\tablenotetext{e}{Temperature (keV).}
\tablenotetext{f}{Neutral hydrogen column density was taken to be
$\log N_\mathrm{H}$ = 18.7 (Jensen \etal 1986).}
\tablenotetext{g}{Emission measure 
($10^{52}  \int n_e n_\mathrm{H}  \hbox{d}V$)  cm$^{-3}$; $n_e$ is the
electron density,   $n_\mathrm{H}$  the  hydrogen  plasma density  and
$\hbox{d}V$ the emitting volume.  Source distance is taken to be 47 pc
(Werner \& Rauch 1997).}
\end{deluxetable}

To improve  the accuracy of $\chi^2$  fitting, the SIS  and GIS energy
channels    containing less that  30  events  were  grouped with their
neighbors.   Channels flagged as  bad   were rejected and those   with
corrected pulse heights of $0.3 > E > 10$ keV were ignored.  Best fits
to the data were determined using  the {\sc apec} thermal plasma model
(Brickhouse  \etal\   2000), absorbed  through a  neutral interstellar
column     with      photo-electric    cross-sections   provided    by
Balucinska-Church \& McCammon  (1992).  Spectra from all  four cameras
were employed together to converge on a single model.

The {\sc apec} code (v1.10)   contained within the current version  of
the spectral fitting package, {\sc xspec} v11.2, does not include high
$n$ transitions  of  Fe {\sc xvii--xix}.    Also there are  systematic
problems in line wavelengths for the set of  Ni lines in this version.
Therefore additional  tables for   Fe {\sc xvii};  $n$ =   6, 7 and  8
$\rightarrow$ 2,  Fe {\sc xviii};  $n$ =  6, 7 $\rightarrow  2$ and Fe
{\sc xix}; $n$ = 6 $\rightarrow$ 2 are included  together with a table
of corrected Ni wavelengths, as  described in Brickhouse \etal (2000).
Since these models are  additive, abundances cannot be  decoupled from
model normalizations in these  two cases. Although these  models avoid
errors in fitting  high-$n$ Fe and Ni,  they cannot be used to measure
abundances.

If all abundances are  allowed to roam  as independent fit parameters,
then a statistically  significant constraint on any individual element
is not possible using this data set.  By trial  and error we find that
significant   detections of   Mg, Fe,   Si,  O and   Ne   (with H-like
transitions at the peak of  the \asca\ response) are  made if we adopt
some reasonable assumptions  concerning   the fractions of the   other
elements contributing to the line spectrum, i.e.,  C, N, Al, S, Ar and
Ca.  Consequently we    describe below four  abundance  models.  Solar
abundance values were taken from Grevesse \& Sauval (1998).

Model 1 assumes  the abundances of all  metals are equal; in this case
the abundance is dominated by the L shell lines of Fe.  Model 2 allows
the abundances of  Mg, Si, O and Ne  to float as free parameters while
the  others are  floating but    assumed equal.   The combined   metal
abundance will again be dominated by Fe. Model 3 is similar to model 2
except that the  Fe abundance is decoupled  from the combined elements
and allowed  to float on  its own.  The remaining  abundances of C, N,
Al, S, Ar and Ca float, but are equal.  Model  4 is identical to model
3 except C,  N, Al, S,  Ar and Ca are fixed  abundances with the  mean
photospheric value of the Hyades  members ([m/H] = $+0.1$; Mart\'{i}n,
Pavlenko and Rebolo 1997, where [m/H] is  the logarithmic ratio of the
element abundance, m, and H abundance, relative to solar).

Realistically, we  expect a whole   range of plasma  temperatures and,
unsurprisingly, single-temperature models   provide poor  fits  to the
data with  $\chi^2 = 571$ for 337  d.o.f.  (model 3).  Two-temperature
models yield statistically significant   fits although they  are still
undoubtedly a  simplification  of  the temperature   structure.  Table
\ref{tab:abundance}   lists  the  best   fit parameters   and $\chi^2$
statistic for each model.  All models  are acceptable according to the
$\chi^2$ statistic so we cannot  adopt a best-model and can generalize
the results only.

Between  Models  2, 3   and   4 we   find general  consistency between
temperatures and emission measures.   If  the neutral hydrogen  column
density was left as a free parameter, in all cases  it was found to be
both consistent with zero and  smaller than the total galactic  column
in  that  direction   (1.58  \dex{21}  cm$^{-2}$;   Dickey \&  Lockman
1990). However, since the \asca\  energy range is  not well suited for
the measurement of the expected column in front of \vtau, we adopt the
value derived from  \exosat\ measurements  by  Jensen \etal  (1986) of
$N_\mathrm{H}$   =  5 \dex{18} cm$^{-2}$.   The  flux from  the source
according to these models is 2.4 \dex{-12} erg\,s$^{-1}$\,cm$^{-2}$.

For each model, Fe abundance is most-likely sub-solar, the opposite to
the     solar   corona, whereas Ne   is     greater  than  solar.   In
Fig.~\ref{fig:spectrum} we  show the best-fit model  3,  folded on the
energy-resolved count rates from  SIS0  and GIS2 and  their residuals.
Although far from conclusive, the high Ne/Fe ratio is suggestive of an
inverse-FIP effect.  For models 2, 3 and  4, the coronal abundances of
the five significantly-constrained elements are plotted against FIP in
Fig.~\ref{fig:fip}.

\section{Discussion}
\label{sec:discussion}

\begin{figure*}
\begin{picture}(100,0)(10,20) 
\put(0,0){\includegraphics{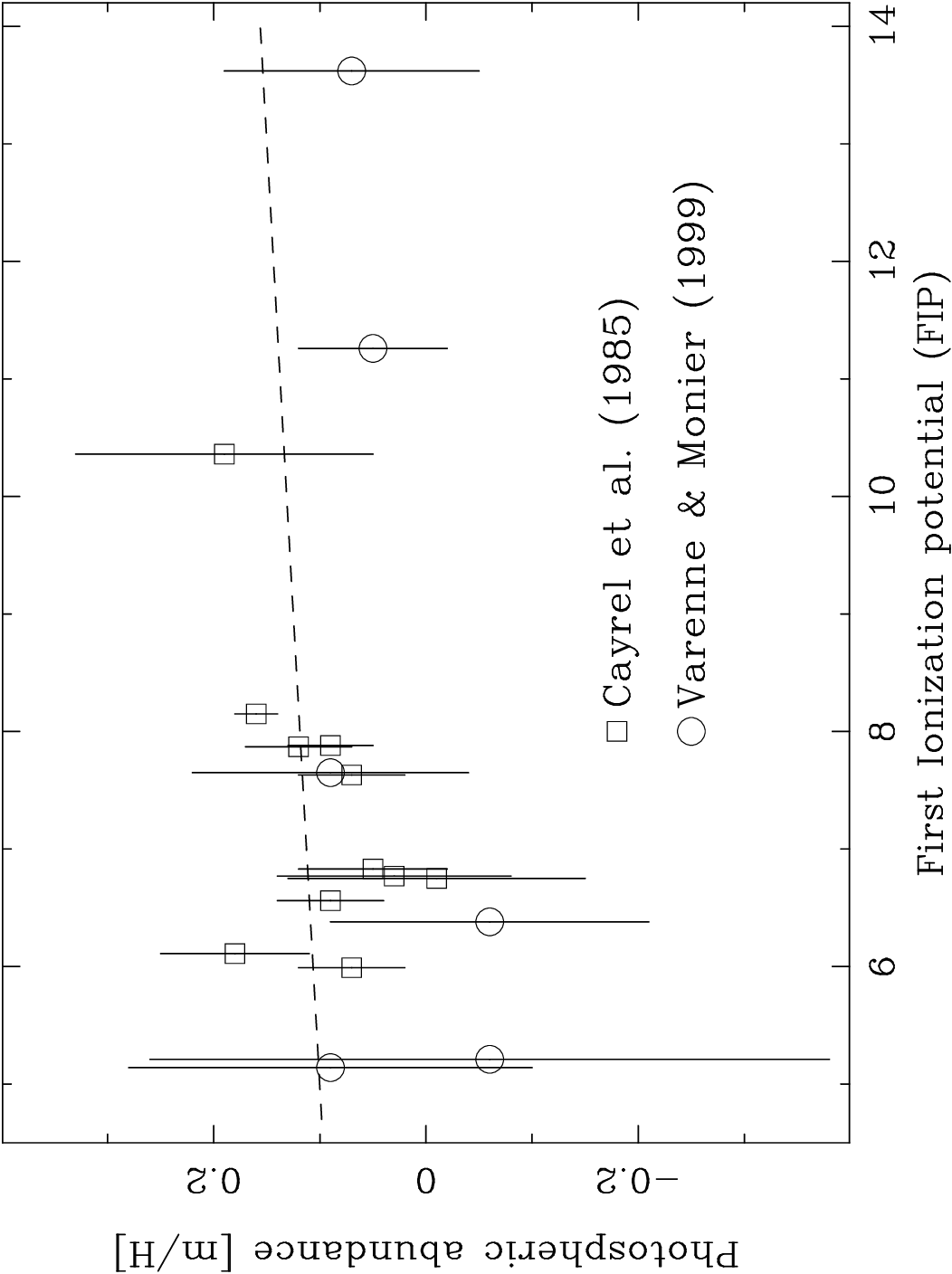}}
\noindent
\end{picture}
\vspace{85mm}
\figcaption[hyades.ps]{Mean photospheric abundances of Hyades dwarfs   
(Cayrel  \etal 1985; Varenne  \& Monier 1999)   plotted as function of
FIP.  The line is the best linear fit to the data. \label{fig:hyades}}
\end{figure*}

For  various   model assumptions,   we   have determined  the  element
abundances of  Mg,  Fe, Si,  O   and Ne in  the  corona  of \vtau\ and
demonstrated a  high  Ne/Fe  ratio which  could be   indicative of the
inverse-FIP effect. In the following section, we ask whether we expect
to find fossil abundances from the post common envelope epoch. We also
discuss  the proposed mechanisms  behind the FIP  effect and, finally,
compare  the   coronal abundances    with the  measured   photospheric
abundances of dwarf stars in the Hyades cluster.

\subsection{Common envelope abundances}
\label{sec:ce}

During the white dwarf progenitor's  red giant phase of evolution, the
two stars shared  a common envelope  and, therefore, there was element
mixing   between  them.  Using    LTE    atmosphere fits  to   optical
spectroscopy of  \vtau,  and  providing NLTE   corrections, Mart\'{i}n
\etal (1997) found photospheric abundances consistent with the Hyades,
[m/H] = +0.1, in Al, Ca,  Fe and Si.  The  one exception was Li, where
[Li/H] = $2.4  \pm 0.3$.  This is greater   than a factor $10^2$  more
abundant than normal Hyades members.   Although Li is usually depleted
in convection layers, Mart\'{i}n \etal\ argue  that the rapid rotation
of the K star maintains the fossil level of Li  after sharing a common
envelope phase with  the Li-rich giant (de  la Reza, Drake \& da Silva
1996).  Unfortunately there are no Li transitions in the \asca\ energy
band to test this independently.

Marks  \&   Sarna (1998)  calculate  the   evolution  of  photospheric
abundances on the  companion star of  cataclysmic variables, including
the effects  of a common  envelope epoch.  They indicate that, indeed,
photospheric abundances in the K star are expected to remain unchanged
through the  common  envelope phase,  and remain  constant  thereafter
until the white dwarf accretion rate is sufficient to drive a cycle of
nova eruptions.  Therefore  we do not  expect to find evidence for the
common envelope phase  in the current  measurements, however, peculiar
abundances  may yet  be present if  the   proposed \iue\ detection  of
expanding nova material around \vtau\ by Bruhweiler \& Sion (1986) and
Sion   \etal\  (1989)    is  confirmed.     However   subsequent \hst\
spectroscopy shows  no indication  of  shell features but  does detect
transient coronal ejection events (Bond  \etal  2001). These may  well
have  been  mistaken for an expanding   shell before coronal ejections
were time-resolved  by \hst.  In  light of  both the observational and
theoretical evidence, we will assume  that the photospheric abundances
of the companion star  in \vtau\ are identical  to the average  Hyades
dwarf. This is useful later for comparing the coronal and photospheric
abundances.

\subsection{FIP bias}
\label{sec:atmospheres}

Observations have indicated  that the coronal, energetic particle  and
cosmic ray  abundances in  the sun are   all different from  the solar
photospheric content (Meyer  1985a,b).  This is probably equally  true
for active  stars (Brinkman \etal 2001;  Audard \etal 2003).  While in
most cases the photospheric abundances  remain uncertain, there is  at
least   one strong piece    of evidence for  fractionation regions  in
between  stellar photospheres and  coronae.  It derives  from the fact
that coronal   abundances are correlated  with  elemental FIP.  In the
solar atmosphere the  ratio  of  coronal  to photospheric   abundances
decreases with FIP.  When comparing stellar  coronal abundances to the
solar photosphere some sources follow this trend while others show the
``inverse FIP effect'', i.e. an  increasing ratio with increasing  FIP
(Brinkman \etal 2001).  The Ne/Fe ratio in  \vtau\ is suggestive of an
inverse-FIP trend.

FIP effects are thought to be directly associated  with the process of
element  fractionation in stellar  atmospheres (H\'{e}noux 1995).  The
ionized fraction of each element will depend on the plasma temperature
in the fractionation region.  Some process is required to decouple the
ionized and   neutral  plasmas.  Possibly  the  charged  particles are
accelerated along field lines  in loops above the stellar  photosphere
(Wang 1996).     Flare ejection provide   another  mechanism to propel
material into the corona  (Schmelz 1993).   The abundances of  coronal
elements as a function of FIP has been observed to change during X-ray
flares.  Audard, G\"{u}del  \& Mewe (2001) find  an inverse FIP affect
in HR  1099 during  quiescence but a  quenching  of this effect during
flares.

\subsection{Abundances in the Hyades}
\label{sec:hyades}

Abundances in most stellar photospheres are not  well known.  Instead,
it is  common practice to  compare stellar coronal abundances with the
solar   photosphere.     Rather  than   adopting    solar photospheric
abundances, ideally we  would  prefer to  have direct  measurements of
photospheric abundances in the K companion of \vtau\  in order to make
an  unbiased comparison between  photosphere   and corona.  These  are
poorly determined    generally because of  uncertainties  in  galactic
element  mixing and the details  of stellar spectral synthesis models.
Fortunately, though,  \vtau\ is a member  of the nearest open cluster,
the Hyades (Werner  \& Rauch 1997),  where photospheric abundances are
considerably better determined than most coronal X-ray sources.

Metallicities in open clusters are indicators of the quantity and rate
of  chemical mixing  in    the  galaxy.  However  stellar    abundance
determinations are  generally plagued by a  wide  range of biases (see
e.g.  Griffin \& Holweger 1989).  Various techniques have consequently
supplied   abundances  for the Hyades   cluster  ranging from [Fe/H] =
$-0.09$ (Tomkin \& Lambert 1978)  to [Fe/H]  = $+0.42$ (Gustafsson  \&
Nissen 1972).   

Although not unequivocally free of  bias, Cayrel, Cayrel de Strobel \&
Campbell (1985) determine a   mean  [Fe/H] enhancement over solar   of
$+0.12 \pm 0.03$ using high dispersion spectroscopy of  a sample of 12
Hyades dwarfs distributed tightly about the solar spectral type.  This
result was  determined by comparison of the  curve of growth  over the
linear and  saturated branches  of Fe{\sc  i} lines  with photospheric
spectral  models.   A large surface  fraction  of active  regions over
these stars   would probably  result   in  an  under-estimate  of  the
abundance, but an identical estimate of [Fe/H] = $+0.14 \pm 0.01$ from
a smaller sample of  Fe{\sc ii}  lines  suggests that any bias  due to
chromospheric activity   is small  among the measured    cases.  While
systematic errors may still be present, this result has been confirmed
with a sample   of F  stars with    $v\sin{i} < 30$   km\,s$^{-1}$  by
Boesgaard (1989), finding [Fe/H] = $+0.13 \pm 0.03$.  Consequently, by
comparison  with  the \asca\  results, Fe abundance  is  higher in the
photosphere compared to the corona of \vtau, within the uncertainties.

Cayrel \etal (1985) provide coronal abundances for Si{\sc i}; [Si/H] =
$+0.16  \pm  0.04$.  Varenne \&  Monier  (1999) present photospheric O
abundances for 26 F stars in the Hyades with an average value of [O/H]
= $+0.07$. In these cases there are no significant differences between
the coronal and  photospheric populations, but the coronal  abundances
are poorly  defined   using the current  data.  However,   as in other
sources (e.g.\  Brinkman \etal 2001), the  suggestion of a coronal FIP
bias in \vtau\ hinges crucially on the measurement of Ne.

Critically, photospheric   Ne  has not  bean  measured in   the Hyades
because  the low  photospheric   temperatures    do not provide     an
environment  for noble gas  emission. We combine the Hyades abundances
from Cayrel \etal and Varenne \& Monier  in Fig.~\ref{fig:hyades}.  We
do not include  the two  stars in the   Cayrel \etal sample  that  are
thought to be  active sources.  Data points  from a sample of  cluster
members have been  averaged and uncertainties  obtained by calculating
the standard deviations for  each species.  A linear least-squares fit
yields [m/H] =  $0.069^{+0.140}_{-0.140} + 0.006^{+0.018}_{-0.017} E$,
where  $E$  is  the FIP  in   eV.  Uncertainties are  the   90 percent
confidence limits, i.e.  the   abundances are consistent with  a  flat
distribution across the range of FIP.

We fit    the      slope of  the   X-ray   FIP    distributions   from
Fig.~\ref{fig:fip}, including  the Ne data.   Clearly the FIP-trend is
consistent   with a constant if  Ne  is ignored.  Least-square fitting
yields slopes consistently of  $0.04 \pm 0.03$ for  models 2, 3 and 4.
Uncertainties are 90 percent confidence limits.  Consequently, with Ne
included, the coronal FIP distribution is steeper than the Hyades mean
photospheric distribution.

\section{Conclusion} 
\label{sec:conclusion}

We have measured model-dependent  coronal abundances of  Mg, Fe, Si, O
and Ne  in  the corona of  the  post-common envelope binary  \vtau.  A
single-temperature plasma   model does not  fit the   data adequately.
Binary evolution  calculations  have predicted that  we   would see no
symptoms  of element mixing  during  the common-envelope  epoch and we
indeed find  no  evidence for it;  i.e.,  abundances are   not unusual
compared  to the coronae of   single stars or   wide binaries.  In all
likelihood, the coronal Fe   abundance is significantly less than  the
Hyades photospheric mean.  This is in direct  contrast to the ratio of
Fe in the  solar corona  which  is overabundant relative to  the solar
photosphere.  There is evidence for an inverse FIP bias, although this
relies entirely on  our measurement of  coronal Ne abundance.  Due  to
the lack   of  low-ionization  Ne lines   in   the  optical  band, the
photospheric abundance of this element  cannot be directly compared to
the coronal value, adding further uncertainty.  While CCD spectroscopy
is  not currently an  ideal  method for abundance determination, these
data indicate that  \vtau\ is a  viable and interesting target for the
grating instruments on-board the \chandra\ and \xmm\ observatories for
such a purpose.

\acknowledgments  
This paper    was based on    data obtained   from  the   High  Energy
Astrophysics  Science Archive Research   Center (HEASARC), provided by
NASA's Goddard  Space Flight  Center.  We  thank  Nancy Brickhouse for
supplying additional Fe and Ni atomic data.

\end{document}